\documentclass{raa}            % referee version: for submission

\usepackage{graphicx,times}             %for PS/EPS graphics inclusion, new
\usepackage{natbib}
\usepackage{amssymb,amsmath}
\usepackage{fontspec}
\usepackage{siunitx}

\bibpunct{(}{)}{;}{a}{}{,}

\usepackage[pagebackref=true]{hyperref}

\begin{document}

  \title{Mock Observations for the CSST Mission: CPI-C--Instrument Simulation}
%   \subtitle{I. Place Your Subtitle Here}

   \volnopage{Vol.0 (20xx) No.0, 000--000}      %%preserved for Editor. DOn't remove!
   \setcounter{page}{1}          %%starting page, preserved for Editor. DOn't remove!

   \author{%% Put your Chinese name in "( )" if you like. Note to open line 11 "\usepackage[UTF8]{ctex}"
         Gang Zhao  \inst{1,2}
    \and Yiming Zhu  \inst{1,2}
    \and Jiangpei Dou \inst{1,2}
    \and Yili Chen  \inst{1,2}
    \and Zhonghua Lv  \inst{1,2}
    \and Bingli Niu  \inst{1,2}
    \and Zhaojun Yan  \inst{3}
    \and Bo Ma  \inst{4,5}
    \and Ran Li \inst{6}
   }

   \institute{Nanjing Institute of Astronomical Optics \& Technology, Chinese Academy of Sciences, Nanjing 210042, China; {\it gzhao@niaot.ac.cn; jpdou@niaot.ac.cn}\\
        \and
            CAS Key Laboratory of Astronomical Optics \& Technology, Nanjing Institute of Astronomical Optics \& Technology, Nanjing 210042,China\\
        \and 
            Shanghai Astronomical Observatory, Chinese Academy of Sciences, Shanghai 200030, China
        \and
            School of Physics and Astronomy, Sun Yat-sen University, Zhuhai, Guangdong 519082, PR China
        \and
            Center of CSST in the great bay area, Sun Yat-sen University, Zhuhai, Guangdong 519082, PR China
        \and 
            School of Physics and Astronomy, Beijing Normal University, Beijing 100875, China
\vs\no
   {\small Received 20xx month day; accepted 20xx month day}}

\abstract{
To support the development of the data processing pipeline and the scientific performance assessment for the Cool Planet Imaging Coronagraph (CPI-C) on the Chinese Space Station Survey Telescope (CSST), we have developed the end-to-end instrument simulation program, CPISM. This paper details the core modules of CPISM that simulate the CPI-C instrument, focusing on the simulation of the high-contrast imaging optical system and the visible-band science camera. We modeled key optical components, such as the transmission apodizing filter, the wavefront corrector, and the focal plane mask using the HCIPy package. A $10^{−8}$ contrast dark hole region, consistent with design specifications, was simulated using the Electric Field Conjugation (EFC) optimization method, and broadband observation effects were considered. For the science camera, which is an electron multiplying charge-coupled device (EMCCD), we established a detailed model encompassing photon collection, charge transfer, electron multiplication (EM), and readout processes, based on test data. This model simulates complex instrumental features including dark current, charge transfer efficiency, clock-induced charge, multiplication noise factor, and various readout effects like striping and drift. We also proposed and validated an improved statistical model for the EM process to enhance simulation efficiency. CPISM can generate simulated images containing rich instrumental details, closely similar to the expected real observational data, thus laying the foundation for the development and verification of CPI-C data processing algorithms and preparations for future scientific research.
}

   \authorrunning{G. Zhao, Y. M. Zhu \& J. P. Dou et al. }            %author_head in even pages
   \titlerunning{CPI-C instrument simulation}  % title_head in odd pages

   \maketitle
%% The author head (on even pages) and the title head (on odd pages) will be
%% automatically extracted from \author{} and \title{}. Whenever the title is too long,
%% you will be asked to supply a shorter one by inserting either \authorrunning{} or
%% \titlerunning{} before \maketitle. Anyway, you can specify your own heads.
%%
%%
%% Note: In the following text body of your manuscript, please note several differences from
%%       other major journals:
%% (1) \subsection{Please Capitalize the First Letter of Each Notional Word in Subsection Title}
%% (2) Please Capitalize the First Letter of Each Notional Word in all tables' captions

%
%________________________________________________ sections below
%
\section{Introduction}           %% first-level sections will be auto-capitalized
\label{sect:intro}

%% Authors can give a citation as 'Michel et al. 1992'.
%% You may also use \cite, \citep and \citet for citation, and use Table~1 or Figure~1
%% and so forth. Using \ref and \label for cross-references of Tables/Figures
%% is a good way in adjusting/adding/removing text, tables or figures.

The \textit{Cool Planets Imaging Coronagraph} \citep[CPI-C\footnote{In Table~\ref{tab:acronyms}, we list all the acronyms in the paper.},][]{dou25} is one of the five back-end instruments of the \textit{Chinese Space Station Survey Telescope} \citep[CSST,][]{csst25}. It is designed for ultra-high contrast detection and multi-band photometry studies of extrasolar planets and circumstellar disks in visible to near-infrared wavelengths. Utilizing pupil apodization \citep{ren07, ren10}, and phase modulation methods \citep{dou16}, CPI-C creates two dark regions with a contrast ratio up to $10^{-8}$ near the center of the star's image. In these dark regions, multi-band images of extrasolar planets can be obtained. Subsequently, spectral fitting can be used to infer their atmospheric spectra and physical properties. The construction of the coronagraph instrument is currently underway, and the development of the data reduction pipeline and preliminary scientific research are also in progress. To support these tasks, we have developed an end-to-end CPI-C image simulation tool - \texttt{CPISM} (CPI-C Image Simulator). \footnote{https://csst-tb.bao.ac.cn/code/csst-sims/csst\_cpic\_sim}. The \texttt{CPISM} is developed using the Python language. When users input target information such as stellar spectral type and planetary radius, as well as instrumental parameters like exposure time and observation bands through the command line, \texttt{CPISM} can generate simulation images that match the format of real data for subsequent processing and analysis.

In modern space astronomy projects, simulation plays an increasingly important role. Simulation tools, integrated with instrument test results, provide a research foundation for target selection, observation planning, and the preparation of observation data processing procedures. The James Webb Space Telescope (JWST) serves as an excellent example. Years before its launch, high-contrast simulation programs, as well as camera simulation programs, point spread function (PSF) simulation programs, and the instrument effect simulation tools were prepared \citep{girard18}. The JWST pipeline has also been developed based on simulation data \citep{gordon22}. The thorough preparation has led to a rapid scientific output. For instance, the direct imaging observation of the exoplanet VHS 1256 b \citep{miles23} proceeded smoothly and rapidly following the observation on July 4, 2022. The complete workflow, including initial data reduction, high-contrast data processing, target photometry, spectral fitting, and determination of planetary physical properties, was accomplished in just two months, with the related paper submitted by September 3, 2022. Similarly, for the CGI instrument of the Roman Space Telescope (under construction), a series of simulation tools were developed \citep{douglas20}, including planetary target transmittance spectrum models, coronagraph high-contrast simulation, and complete end-to-end simulation. These simulation tools have been used to validate data processing algorithms \citep{ygouf2021} and determine exoplanet parameters \citep{turnbull21}.

As an end-to-end simulation tool, \texttt{CPISM} consists of an observation target simulation module, an observation effect simulation module, a high-contrast imaging simulation module, and a camera simulation module.
%As a part of the CSST ground-based science data processing system, 
The core objective of the \texttt{CPISM} is to provide test images for the CPI-C data reduction pipeline. The pipeline primarily focuses on instrumental effects correction. It takes the level 0 data (raw data) from CPI-C as input, and processes this data to correct detector effects and remove background light and cosmic ray influences, generating level 1 image data. To meet the needs of processing detector effects, \texttt{CPISM} thoroughly models the EMCCD camera, considering the results of camera qualification tests. The simulation covers various stages of the camera's operation, including photon collection, vertical transfer, EM amplification, and readout. To address the needs of cosmic ray processing, the \texttt{CPISM} constructs a cosmic ray model suitable for the CPI-C's camera, based on the characteristics of cosmic rays from Hubble Space Telescope observations \citep{heyer04, miles21}. Currently, configuration testing of the CPI-C data reduction pipeline is being carried out based on data generated by \texttt{CPISM}. Furthermore, \texttt{CPISM} is also designed to validate CPI-C science \citep[e.g. ][]{bao25}. To this end, the \texttt{CPISM} semi-quantitatively simulates  high-contrast imaging dark holes with a contrast ratio of $1\times
10^{-8}$, as well as the reflected spectra of exoplanets with different metallicities and cloud characteristics.

The \texttt{CPISM} will be introduced in two papers. In the first paper of the series, \cite{zhu24} provided an overall introduction to the \texttt{CPISM}, as well as the details of simulation of observation targets and the use of simulated data for scientific research. In this paper, we introduce the simulation of the CPI-C instrument, focusing on the high-contrast imaging module and the visible band camera. The article is organized as follows: In Section 2, we present the numerical models of the core components of the high-contrast imaging simulation module, including the apodized pupil filter, high-order wavefront corrector, and focal plane mask. We explore methods to optimize the high-contrast dark zone, and achieve a PSF similar to the instrument measurement results. In Section 3, we introduce the model of the EMCCD serving as the visible band scientific camera of CPI-C, and provide a detailed introduction to the modeling of the core process of electron multiplication.  In Section 4, we summarize the paper.

\section{Simulation of optics of High Contrast imaging}
\begin{figure}
\centering
\includegraphics[width=9cm]{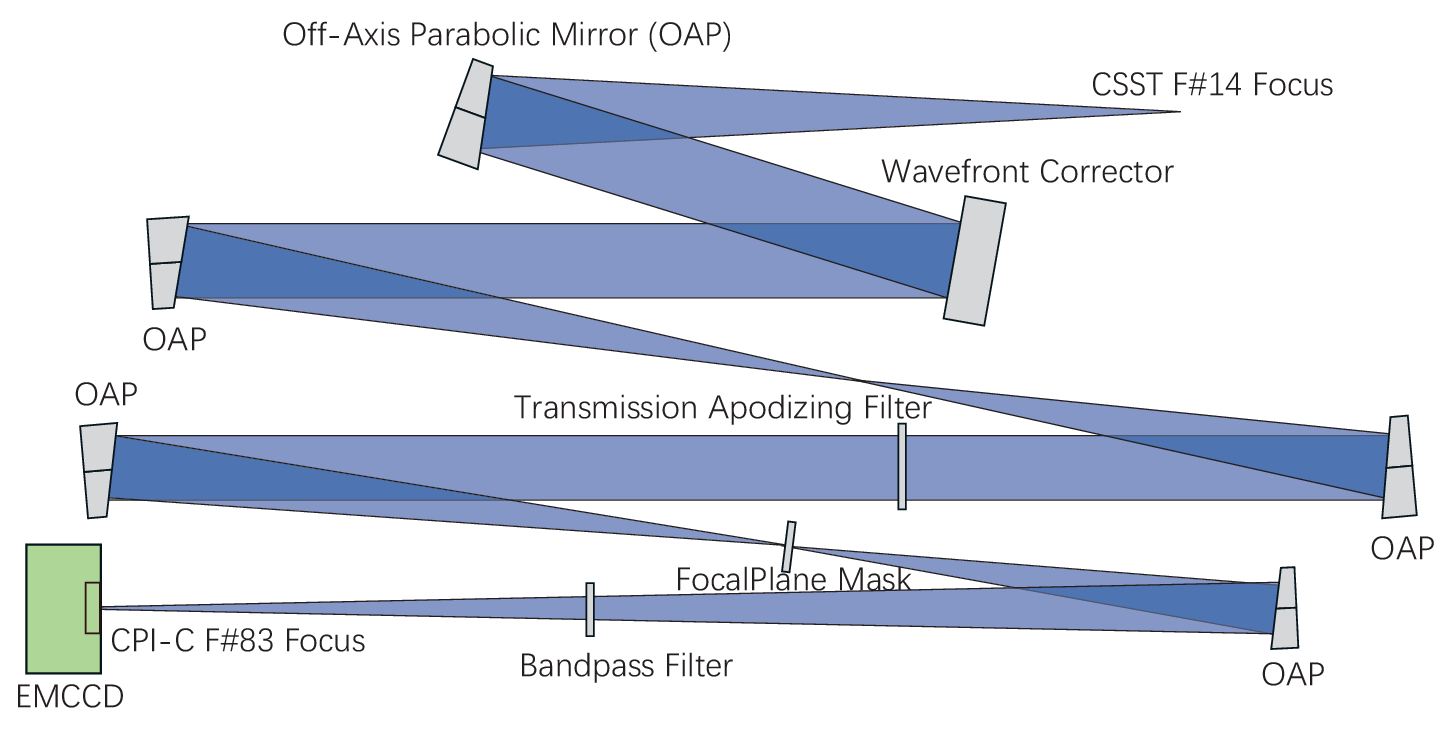}
\caption{The schematic of the CPI-C optical layout.}
\label{layout}
\end{figure}

CPI-C achieves high-contrast imaging of $10^{-8}$ through its core components: the wavefront corrector (which is a deformable mirror), and the focal plane mask. The schematic of its optical layout is shown in Figure \ref{layout}.\footnote{This schematic is a simplification of the actual optical path; the real optical path also includes tip-tilt correction mirrors, reflectors, beam splitters, as well as wavefront sensing branch and near-infrared imaging branch, which are not depicted in the schematic.} The wavefront corrector and the apodizing filter are located at the conjugate position of the pupil plane in the optical path, modulating the amplitude and phase of the wavefront; the focal plane mask is at the secondary focal plane of the optical system, blocking the starlight and the cross-shaped diffraction structures. After passing through the secondary focal plane, the light is focused by an off-axis parabolic mirror and finally imaged on the visible light imaging camera at the F\#83 focus. 

We use the Python-based HCIPy package \citep{por18} to model and simulate high-contrast devices and the optical propagation. During the modeling process, various device parameters, such as the spot size at the DM and the number of DM actuators, are based on design values. However, details such as the aberrations in the system and the DM surface shape when creating the dark zone are not based on actual measurements from the actual optics, which means that the simulated PSF does not completely match the PSF of the real instrument but has similar shapes, and parameters, such as contrast and inner working angle.

\subsection{Modeling the Core Components of High Contrast Imaging}
\subsubsection{The Transmission Apodizing Filter}
\begin{figure}
\centering
\includegraphics[width=14cm]{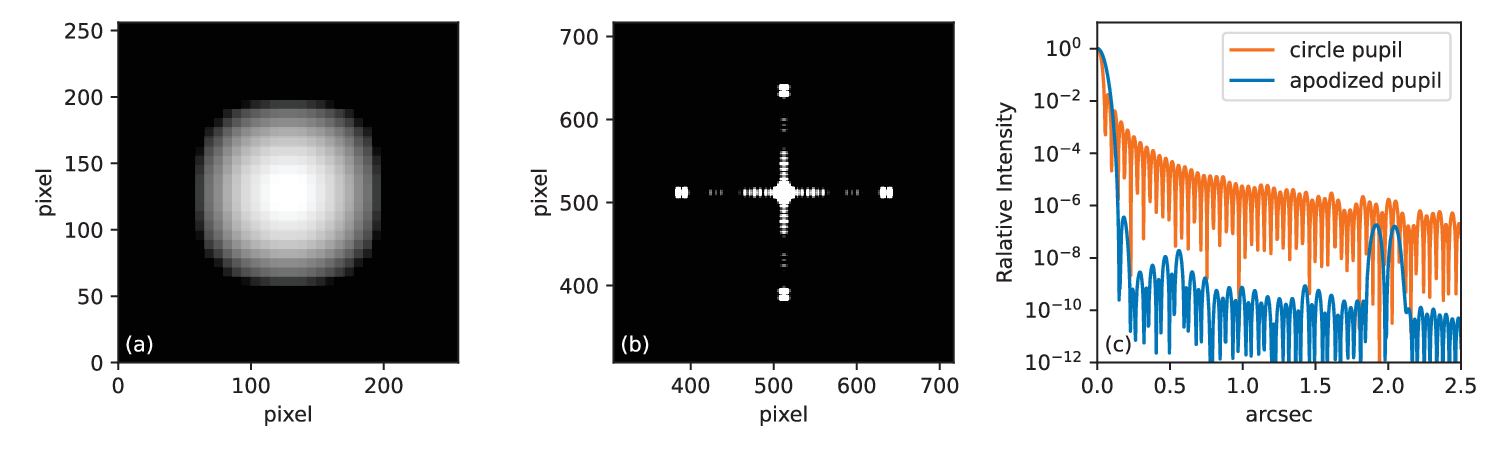}
\caption{Model and simulation results of the Apodizing Filter. a) Transmittance distribution of the Apodizing Filter, with white representing higher transmittance. b) The theoretical point spread function after pupil apodization at 662nm. Photon noise and Readout noise are not considered in the image c) The contrast curves after apodization (blue line) compared to that of an apodized circular aperture. The contrast curve is along the square diagonal direction.}
\label{APD}
\end{figure}

When the light of a point source passes through the telescope, the bright diffracted light can drown out the faint planetary light, making it difficult to directly image the planet. 
The transmission apodizing filter located at the conjugate pupil plane of the telescope, by modulating the electric field intensity on the telescope's pupil, can alter the energy distribution of the PSF, and suppress the diffraction from the telescope aperture. 
The transmission apodizing filter is designed based on the principle of \cite{ren07}, and is composed of 32 strips with different transmittance in both the horizontal and vertical directions. As shown in Figure \ref{APD}a, the central strip has a higher transmittance, with lower transmittance on the sides. 
In the \texttt{CPISM}, we construct it using the \textit{Field} class from the HCIPy module, and the transmittance of the apodizing filter can be configured through the \textit{apm.fits} file in the \textit{refdata} directory. The filter model used in simulation is based on the design values and includes random biases to reflect the influences of manufacturing errors. 
Figures \ref{APD}b and \ref{APD}c show the theoretical apodized PSF image at \SI{662}{nm} and the contrast curve, respectively. 
It can be seen that the filter, while increasing the size of the central bright spot and the cross structure, can effectively suppress diffracted light, achieving a contrast below $10^{-8}$. 
Additionally, multiple orders of diffraction spots can be observed in the image, which can be used for subsequent calibration of target brightness and is an advantage of a step transmission filter.

\subsubsection{The Wavefront Corrector}
In CPI-C's optical path, a 952-actuator Micro-electromechanical Systems (MEMS) deformable mirror (DM) is used as a wavefront corrector. It has two purposes. Firstly, it is used to correct residual aberrations in the system, which may be introduced by the optical path of the telescope or caused by temperature changes and other quasi-static low-order aberrations within the system. Secondly, the wavefront corrector can form a specific surface shape to create a high-contrast imaging dark zone at the focal plane, for details, see \cite{dou16}. We model the DM using the \textit{DeformableMirror} module in the HCIPy package. Since this is a semi-quantitative simulation, we have not measured the actual DM response function but used the built-in DM response function provided by the HCIPy package.

In our simulations, we did not model in detail the optical system of the telescope, such as the imaging quality and tracking accuracy, because the CPI-C integrates wavefront correction and tracking subsystems. These subsystems are designed to compensate for telescope aberrations and tracking errors. During the closed-loop correction process for high-order wavefront aberrations, control accuracy is mainly limited by the precision of aberration detection and control. In the semi-quantitative simulation, which did not model the closed-loop correction process, we introduce a random offset to the DM control voltage to represent control errors. The magnitude of the random offset can be set via parameters.

\subsection{High Contrast Optimization}
The aberrations can significantly degrade image quality as well as the contrast in the CPI-C system. Aberrations are induced by various sources, including the manufacturing process of optical components, thermal changes and gravitational effects. Under the influence of aberrations, the PSF corresponding to the apodizing filter is estimated to only achieve a contrast of $10^{-6}$. For the CPI-C instrument, a high contrast calibration mode has been designed. In this mode,  Stochastic Parallel Gradient Descent (SPGD) algorithm \citep{dou16, dong11} is employed to correct the aberrations, and optimize high-contrast imaging dark zones of $10^{-8}$ at the focal plane. The dark zones are square-shaped, located on one side of the Point Spread Function (PSF) core. Each square region spans the angular separation range from 3 to 16 $\lambda/D$ under theoretical optimal performance. For observations at 662 nm, this corresponds to a physical range of approximately 0.27$''$ to 1.0$''$ from the central star.

In our semi-quantitative simulation, to obtain the optimized shape of the Wavefront Corrector, we did not use the SPGD algorithm adopted by the instrument but used a more efficient Electric Field Conjugation (EFC)  method \citep{matthews17a} for optimization, to form dark holes similar to the design value. The exploration of the EFC method in pupil apodization high-contrast imaging systems may provide a reference for the design of next generation of CPI-C instrument.

\begin{figure}
\centering
\includegraphics[width=14cm]{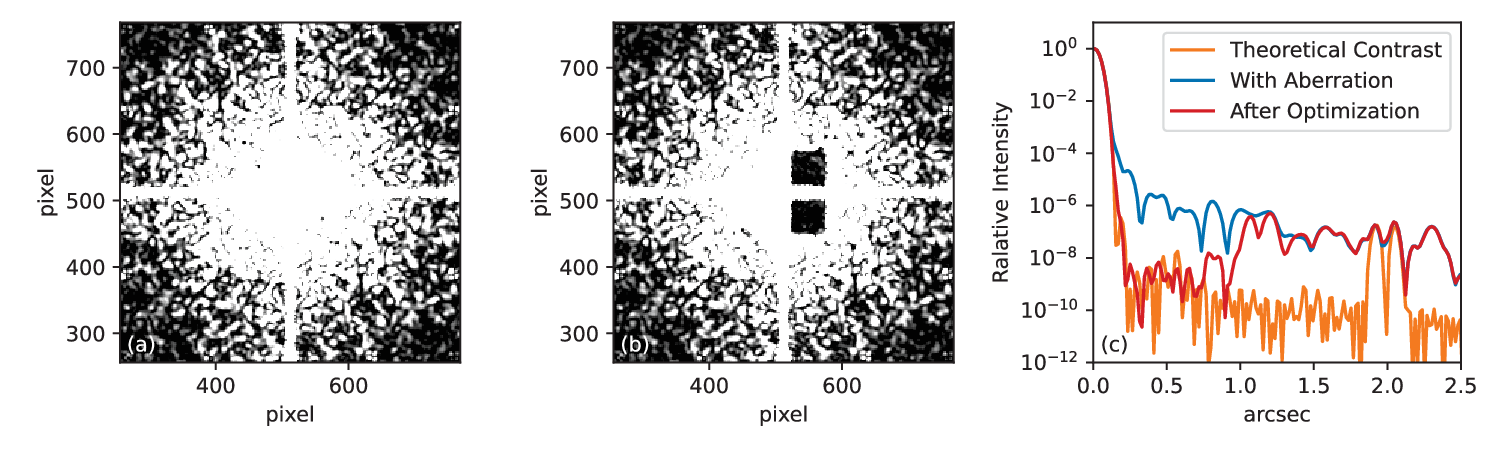}
\caption{(a) The PSF with aberrations at the first focal plane of CPI-C at 662nm. Photon noise and readout noise are not considered.  (b) $10^{-8}$ dark zone after high contrast optimization at 662nm. (c) The contrast curves. Orange line indicates the theoretical curve of the system. Blue line indicates the case with aberrations, and Red line the contrast curve after optimization. The contrast curve is along the square diagonal direction.}
\label{contrast}
\end{figure}

Before optimization, it is necessary to model the system's aberrations. We conducted a qualitative simulation of the aberrations, which have phase and amplitude components when they propagate to the pupil plane.  Under the influence of these aberrations, speckles appear around the PSF (see the example PSF of 662nm in Figure \ref{contrast}a). The resulting contrast, indicated by the blue solid line in Figure \ref{contrast}c, is about $10^{-6}$. This value is on the same order of magnitude as the instrument design specification.

The details of the process of EFC can be found in the Appendix A. After EFC correction, the final optimized imaging dark zone is shown in Figure \ref{contrast}b at 662nm, and the contrast curve is shown by the red solid line in Figure \ref{contrast}c. It can be seen that the final contrast of the dark zone is $10^{-8}$, consistent with the instrument's design. Compared to the contrast curve of the PSF without wavefront aberrations (orange solid line in Figure \ref{contrast}c), the optimization process successfully corrected the speckles in the dark holes, achieving contrast levels similar to the aberration-free PSF. 

The DM primarily influences the inner part of the PSF ($<1.2''$, low spatial frequency region), while leaving the outer part (high spatial frequency region) largely unchanged. This limitation is due to the Nyquist sampling theorem; for a DM with 32 actuators across the pupil diameter, it can effectively control aberrations only within a radius of approximately 16$\lambda/D$ in the image plane, thus defining the system's high-contrast outer working angle. During the dark zone optimization process, by defining specific optimization regions, we can generate dark zones located on the same side of the PSF, such as the right side or the top. When running CPISM, dark zones in different orientations can be selected by setting the 、\texttt{dm\_volt\_index} parameter in the configuration file. However, it is not able to achieve centrosymmetric dark zones. This limitation is primarily because the CPIC system has only a single DM, which cannot provide simultaneous correction of both amplitude and phase aberrations in the system. To form centrosymmetric dark zones, at least two DMs are required \citep{pueyo09}. For instance, the CGI instrument on the Roman Space Telescope utilizes two DMs, enabling the formation of centrosymmetric annular or fan-shaped dark zones \citep{trauger16}.

In addition, there are undulating speckle noise within the dark zone, which is also consistent with the measurement. Subsequent high-contrast algorithms such as Angular Differential Imaging \citep[ADI,][]{marois06},  Locally Optimized Combination of Images \citep[LOCI,][]{lafreniere07}, or Optimized Image Rotation and Subtraction \citep[OIRS,][]{dou15} can be used to suppress the speckles. The simulated images generated by \texttt{CPISM} can also provide test data for developing these speckle-reduction algorithms for CPI-C. 

\begin{figure}
\centering
\includegraphics[width=12cm]{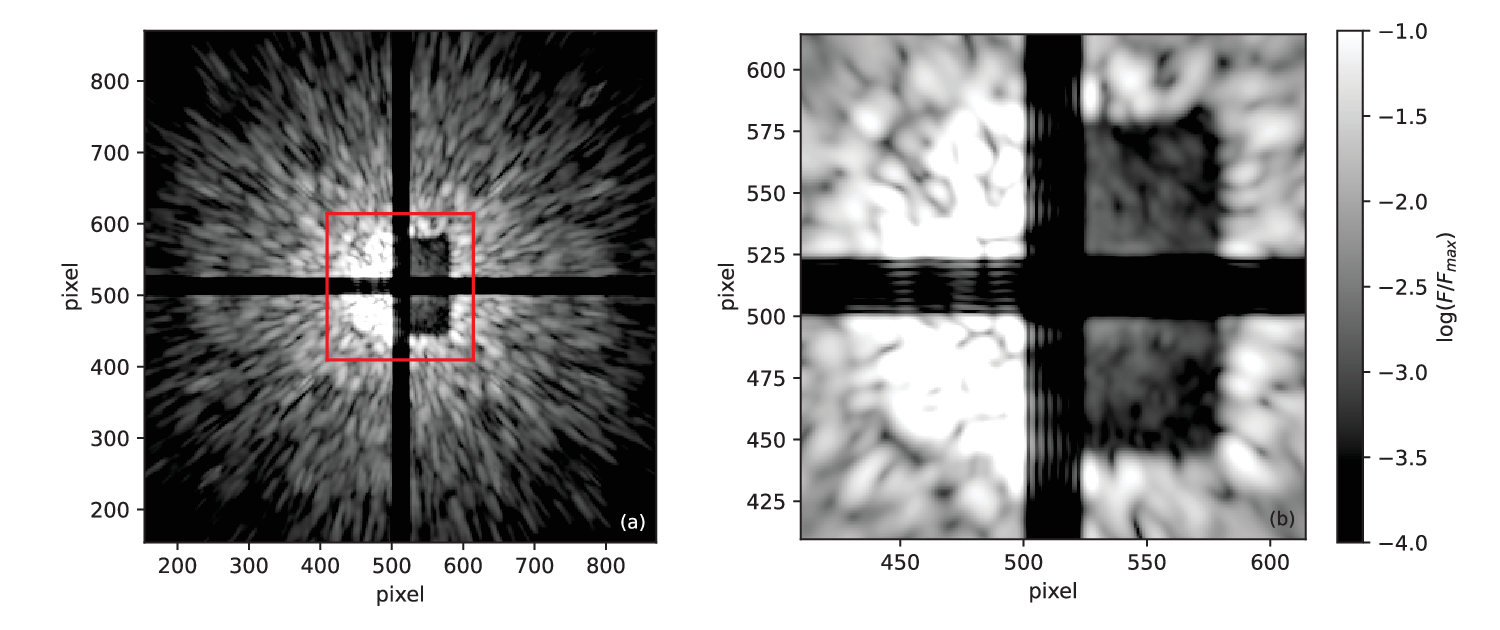}
\caption{Final focal plane image of the CPI-C system for a G0V star at the F662 band . Panel (a) displays the entire focal plane, while Panel (b) presents a detailed zoom-in view of the central region, as indicated by the red square in Panel (a). Note that the dark hole is based on simulations, and it will be calibrated to align with the future ground and on-orbit test results.}
\label{mask}
\end{figure}

\subsection{Focal Plane Mask}
From Figure \ref{contrast}, it can be seen that there is a cross-shaped diffraction structure in addition to the star. This structure is caused by the vertically stacked bands in the transmission modulation filter. 
During visible light imaging, both the star and this diffraction structure can easily saturate and cause overflow. To prevent overflow, a custom focal plane mask is equipped to shield the light of the star and its cross-shaped diffraction pattern. The focal plane mask was fabricated by coating chromium metal film on a fused silica substrate, creating opaque cross arms with a width corresponding to \SI{0.4}{arcsecond}, thereby shielding light within a half-width of $3\lambda/D$ (approximately 0.2 arcseconds) at the F662 band. The throughput in the masked region measures approximately $10^{-6}$. Since the focal plane scale is \SI{1.615e-2}{\arcsecond/pix}, the masked cross-shaped region spans $\sim\SI{24}{pix}$ in the final image.

Our simulation of the focal plane mask and its diffraction effects follows the principle of \citet{dou16}. Using HCIPy's \texttt{ComplexSurfaceApodizer} module, we modeled the mask. For the light through the mask, the wavefront amplitude will be reduced by $10^{-6}$ in the blocked areas. We then calculated the pupil plane wavefront using HCIPy's \texttt{FraunhoferPropagator}. At this pupil plane, high-frequency components induced from the mask were blocked, to simulate the finite aperture of the imaging lens. Light then propagated to the secondary focal plane where science cameras capture images. Figure \ref{mask} shows the result. It can be observed that the star and the cross-shaped diffraction are effectively blocked, leaving only the mask's own diffraction pattern in the obscured regions. 

% To simulate the diffraction effect of this module, we also need to introduce subsequent pupil plane and secondary focal plane simulations in the optical path simulation process. After the light is blocked by the mask at the focal plane, due to the blocking at the exit pupil,

\subsection{Observation Band}
\begin{figure}
\centering
\includegraphics[width=8cm]{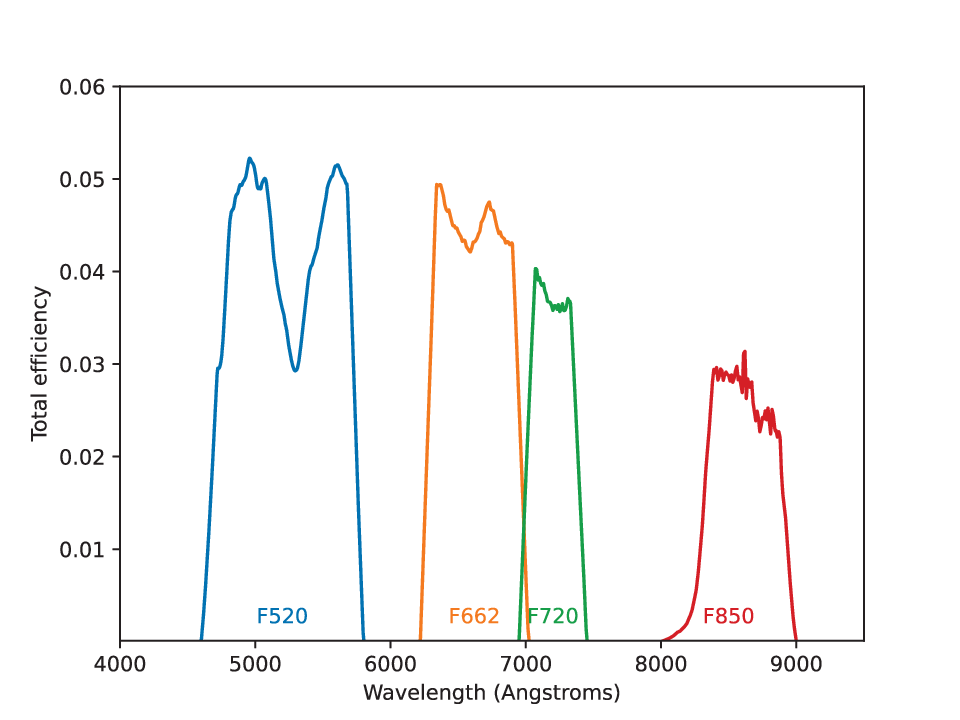}
\caption{The total efficiency curve of the four visible light bands of CPI-C.}
\label{band}
\end{figure}

CPI-C is utilized to conduct observations across four visible light bands (F520, F662, F720, and F850). Figure \ref{band} displays the overall efficiency curve of each band, considering the designed transmission curve of the filters, the measured throughput of the other optical elements in the system and the theoretical quantum efficiency of the camera. The selection of the four-band system is described in detail in a companion study (Zhu et al 2025, submitted). Given the wavelength range of approximately 100 nm for each band, it is necessary to account for the effects of broadband filtering in our simulation program. To this end, each observation band is divided into $M$ narrower sub-bands, where $M$ is an adjustable parameter. For each narrow sub-band, we generate monochromatic images based on the central wavelength position $\lambda_i$, where $i$ ranges from 1 to $M$.

Let $I(\lambda_i)$ represent the monochromatic image observed at the wavelength $\lambda_i$. For each narrow band, we calculate the spectral energy distribution (SED) within the band based on the target's spectral curve $ S(\lambda) $ and the instrument's efficiency curve $T(\lambda)$. Mathematically, for each narrow band, the SED can be expressed as:
\begin{equation}
    SED_i = \int_{\lambda_{i-1/2}}^{\lambda_{i+1/2}} S(\lambda) T(\lambda) \, d\lambda
\end{equation}

where $\lambda_{i-1/2}$ and $ \lambda_{i+1/2}$ denote the lower and upper wavelength boundaries of the $i$-th narrow band, respectively.

Thereafter, we employ the calculated SED values to weight and superpose the corresponding monochromatic images, thereby generating a broadband image. The weighting and superposition process is represented as:
\begin{equation}
    I_{\text{broad}} = \sum_{i=1}^{M} w_i \cdot I(\lambda_i)
\end{equation}

Here, $ w_i $ is the weight of the $ i $-th band, determined by the SED value of that band:
\begin{equation}
w_i = \frac{SED_i}{\sum_{j=1}^{M} SED_j}
\end{equation}

By employing this method, we are able to generate images that take into account the effects of broadband filtering. In Figure \ref{mask}, the image of a G0V star at F662 band is shown. It can be observed that the speckle pattern is elongated into a linear shape. In subsequent analysis, we may be able to utilize such speckle structures to analyze the spectrum of the star.

\section{Simulation of the visible band camera}
\label{sect:emccd}
The CPI-C module uses an EMCCD (Electron Multiplying Charge-Coupled Device) as the imaging detector for the visible focal plane. EMCCDs are widely applied in the observation of faint targets and space astronomy \citep{tulloch11}, such as the CGI module of the Roman Space Telescope \citep{daigle22}. The EMCCD employs electron multiplication technology, which effectively enhances detection efficiency and reduces the equivalent read noise. Before being read out, electrons pass through a series of electron multiplication registers (the number is 604 for the EMCCD used in CPI-C), which contain high-voltage potential wells. When electrons fall into these wells, they have a certain probability of causing an avalanche effect, thereby amplifying the collected signal. In an ideal scenario, if the electron multiplication gain is denoted as $g_A$, the equivalent read noise can be reduced to the level of $1/g_A$.

\begin{figure}
\centering
\includegraphics[width=10cm]{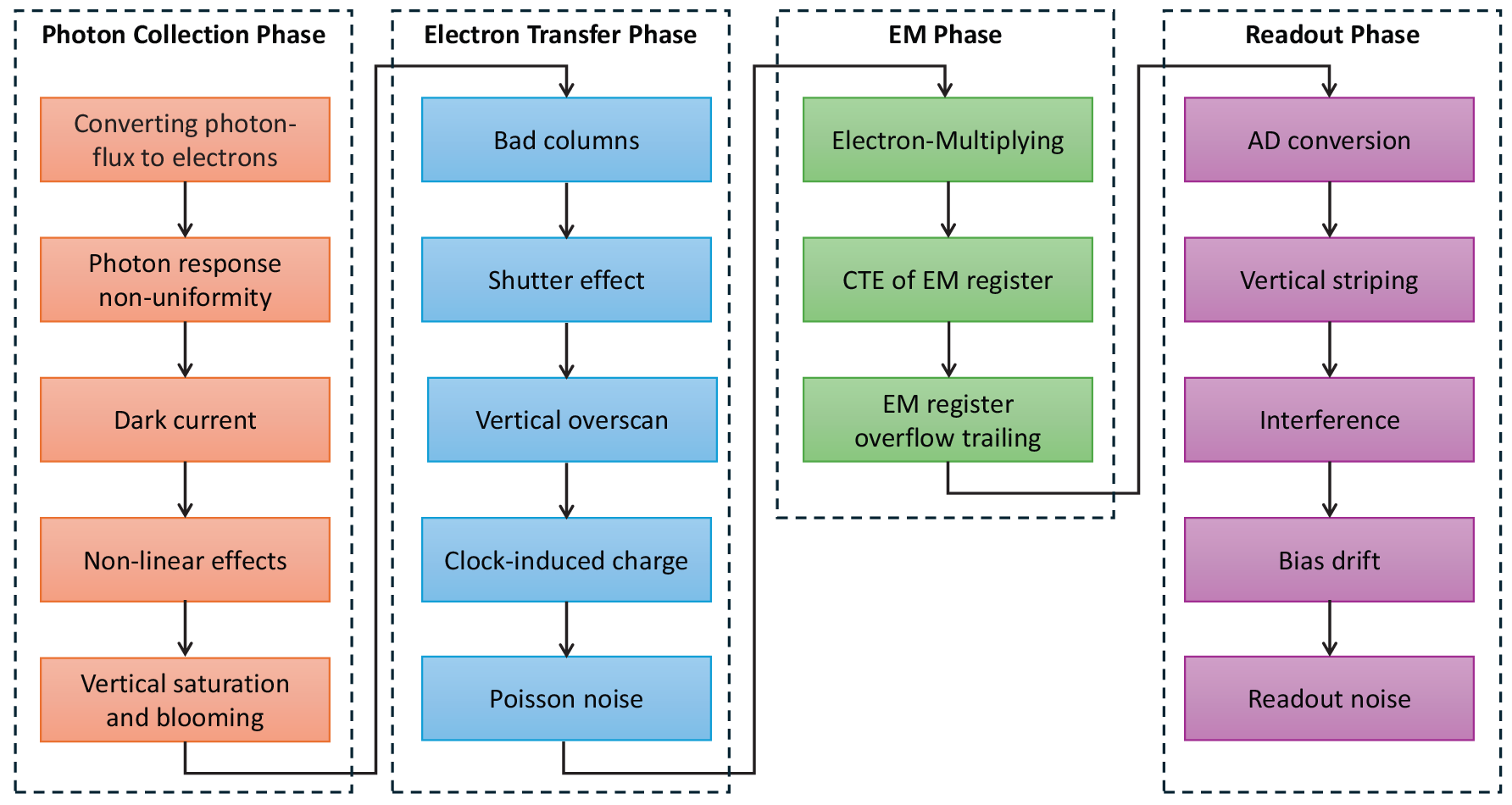}
\caption{The flowchart of the EMCCD simulation process.}
\label{flowchart}
\end{figure}

In the CPI-C simulation program, the EMCCD is modeled, including the photon collection phase, electron transfer phase, multiplication phase, and readout phase. At each stage, various instrumental effects are taken into account to ensure that the final simulated image closely resembles the real image. The flowchart of the entire simulation process is shown in Figure \ref{flowchart}.

\subsection{Photon Collection Phase Modeling}

The collection phase refers to the process where the camera receives photons, converts them into electrons, and accumulates them in the potential wells of the collection area. In the CPI-C simulation program, the photon collection model procedure is as follows:

\begin{enumerate}
    \item First, based on the instrument's throughput curve and the chip's quantum efficiency curve, the 1024 $\times$ 1024 focal plane image, generated by the high-contrast imaging module with units of photons, is converted into an image with units of electrons.
    \item Utilizing flat-field reference data from the configuration file, the spatial non-uniformity of the photoelectric response across the chip's pixels is accounted for.
    \item According to dark-field reference data from the configuration file, dark current and hot pixels are added to the image. The camera has 16 dark reference columns on each side of the image area, 6 rows of dark reference region on the top of the image area, as well as 2 rows on the bottom. The dark reference area and transition region have the same structure as the imaging area, but with an aluminum coating that shields the light. As its name suggests, the dark reference region can be used to estimate the level of dark current of the chip. The image region and dark regions have the total size of 1056 $\times$ 1032. 
    \item A non-linear response to light intensity is added. Currently, in the simulation program, we apply a uniform non-linearity coefficient to the entire image, not considering variations between different pixels.
    \item When the number of collected electrons exceeds the full well capacity, the overflowing charge will spill vertically.
\end{enumerate}

\subsection{Image Transfer Process Modeling}

The EMCCD chip used in the CPI-C requires three vertical transfer operations to capture an image. The first step involves an initial overall transfer at the start of exposure to clear the electrons from the collection area. The second step occurs when the set exposure time has elapsed; the upper half of the chip's collection area transfers the accumulated electrons vertically to the lower half's storage area, which is structurally similar but with an aluminum coating that shields the light. The third step is the line-by-line transfer and readout. These steps enable precise control over the exposure time of each pixel, achieving what is known as an electronic shutter. The simulation of the transfer phase includes the following effects:

\begin{enumerate}
    \item \textbf{Bad Column Effect}: When there are defective pixels in the collection or transfer area that affect the transfer process, this impacts all or part of the pixels in that column, manifesting as a vertical column of abnormal points in the image. In the simulation, the position of bad pixels can be configured to simulate bad columns.
    \item \textbf{Shutter Effect}: As the image transfers from the collection area to the storage area, photoelectric conversion in the collection area continues, leaving a vertical trail in the image. This effect, also known as the smear effect, is related to the transfer speed and is particularly noticeable under strong light and short exposure. The transfer frequency is 500 kHz, effectively providing a 2 ms (1088 row / 500 kHz) short exposure for each position in that column.
    \item \textbf{Overscan}: After the entire image has been fully transferred during the readout process, an overscan operation is performed, involving an additional 18 transfers.
    \item \textbf{Clock Induced Charge (CIC) Effect}: The CIC effect is a particularly important source of noise in EMCCDs. During the readout transfer process of a CCD, there is an extremely small probability that an extra electron will be generated in the potential well, also called a CIC event. For traditional CCDs, this effect is much smaller than the read noise and is generally not considered separately. However, for EMCCDs, the CIC charge will be amplified by subsequent multiplication registers, becoming a significant factor affecting the equivalent read noise of EMCCDs, making it particularly important in our EMCCD modeling. In the CPI-C simulation program, CIC charge is about 0.2$e^-$/pixel/frame, and the pattern can be configured by a reference image.
    \item \textbf{Poisson Randomness}: The electrons from the photoelectric collection process as well as dark current and CIC all follow Poisson distributions. Due to the additive property of the Poisson distribution, $\text{Poisson}(x+y) = \text{Poisson}(x) + \text{Poisson}(y)$, during the transfer phase, we uniformly add Poisson noise to all electrons.
\end{enumerate}

\subsection{Electron Multiplication Stage Modeling}

\subsubsection{EM register modeling}

The CPI-C simulation program includes a section of simulation code for the EM multiplication process based on the gain register model. This code, which is an adaptation of the IDL code from \cite{tulloch10} into a Python version, accounts for both EM multiplication effects and charge transfer inefficiency (CTI) effects. The specific process is as follows:

\begin{enumerate}
    \item For the total EM gain $g_A$, the multiplication coefficient for each gain register is calculated as $g_i = \exp\left(\ln(g_A)/{N}\right)$, where $N=604$ is the count of EM registers .
    \item The electron multiplication process is considered as a Bernoulli process, where each electron has the same probability $p = g_i - 1$ of generating additional electrons, and the probability for each electron is independent. For $m$ electrons, the number of electrons after multiplication by an EM register follows the Binomial distribution $1 + B(m, p)$.
    \item Let the charge transfer inefficiency (CTI) be $q$. After each transfer, the remaining electrons in the register also follow a binomial distribution $B(m, q)$.
    \item By modeling the 604 registers, one can obtain the number of electrons within a pixel after multiplication. Modeling all 1024 $\times$ 1024 pixels of the chip allows one to obtain the complete image after multiplication.
\end{enumerate}

\subsubsection{Statistic Model of Electron Multiplication}
Simulation based on the gain register model is very time-consuming. For instance, simulating a set of one hundred bias frames takes more than 24 hours. To improve the efficiency of the simulation, the \texttt{CPISM} also includes a set of simulation methods based on the statistical distribution of the complete multiplication process. The probability distribution of the output $n$, for an input of $m$ (integer) electrons in an EMCCD has been studied. In \cite{basden03}, this distribution is considered approximately as a Gamma distribution, and \cite{tubbs03} provided a more general form of this distribution as:
\begin{equation}
     P(n; m, g_A) = \frac{ \left(n - m + 1\right)^{m-1}}{(m-1)!  (g_A-1+1/m)^{m} } \exp\left(-\frac{n - m + 1}{g_A-1+1/m}\right),
\end{equation}
By comparing with the simulation results of the complete gain register model, it is found that these two distributions fit well under high gain conditions, but have a significant difference under low gain conditions. To correct the difference at low gains, we introduced a correction factor $c$ to the Gamma distribution, and express the form of this distribution as:
\begin{equation}
     P(n; m, g_A) = \frac{ \left(n - m_0\right)^{m-1}}{(m-1)!  (c g_A)^{m} } \exp\left(-\frac{n - m_0}{c g_A}\right),
     \label{my_dist}
\end{equation}
where $c$ is the correction factor
\begin{equation}
    c=\left[\left(1-g_A^{-1}\right)\left(2g_A^{-1/N}-1\right) \right]^{1/2},
\end{equation}
$m_0=(1-c)g_Am$, and $N$ is the number of EM registers.

\begin{figure}
\centering
\includegraphics[width=10cm]{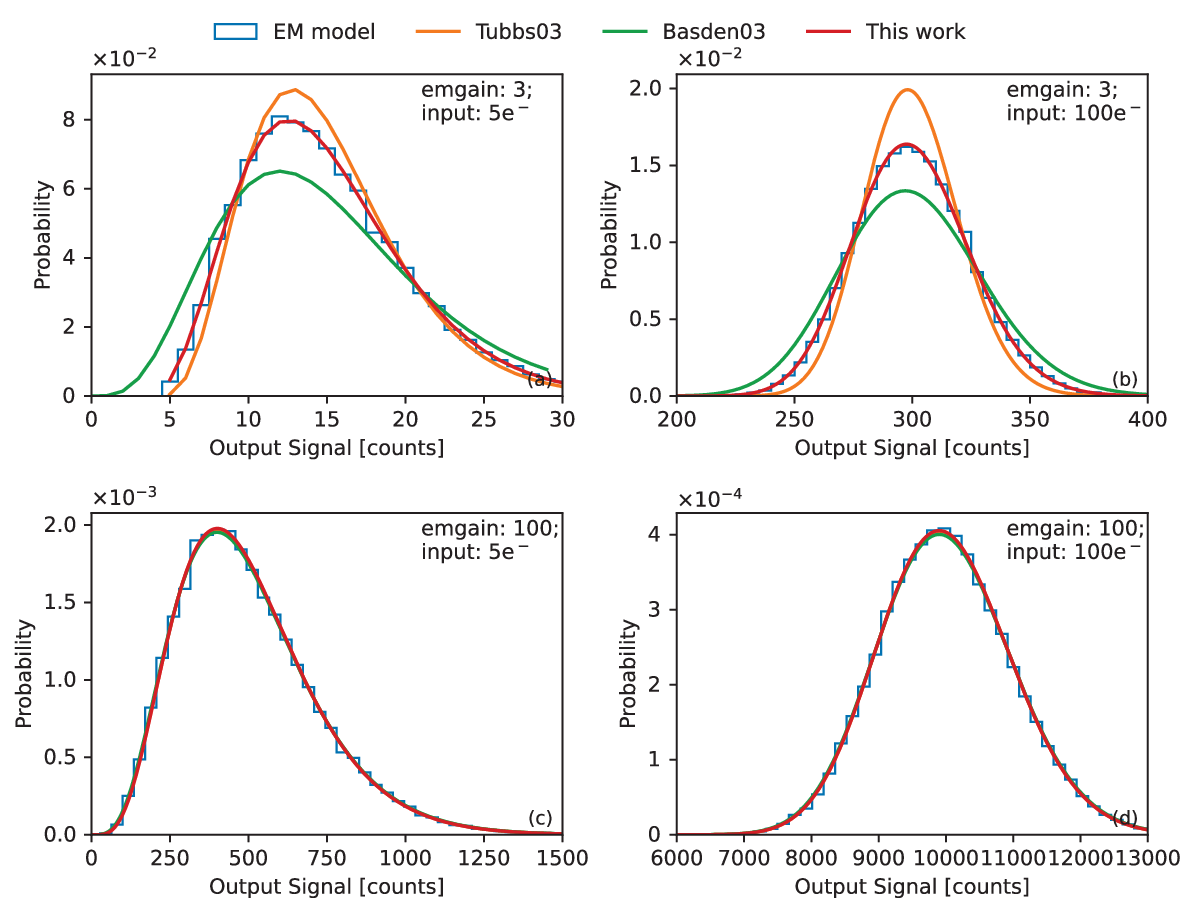}
\caption{Distribution of the output signal of electron multiplication in the EMCCD. The red line represents the theoretical distribution as described by Equation \ref{my_dist}. The orange and green lines correspond to the distributions reported by Tubbs (2003) and Basden (2003), respectively. The blue step-type histogram illustrates the statistical results from the Monte Carlo simulation of the electron multiplication registers.}
\label{distribution}
\end{figure}

In Figure \ref{distribution}, we plot the distribution of output electrons for an input of $m=100$ and $m=5$ electrons at different EM gains. It can be observed that the output distribution is close to a normal distribution curve, and the center of the distribution is located approximately at $g_A \cdot m$ \footnote{For the distribution of Equation \ref{distribution}, the mean value is $g_A m$, while the mode (the peak position where the probability reaches maximum) is $g_A (m - c)$, indicating a left shift of the distribution center relative to the mean value.}. For higher gain ($g_A$=100) situations, all three theoretical distributions are similar to the distribution provided by the complete register simulation. However, for low gain cases such as $g_A=3$, our formula (Equation \ref{my_dist}) can better reflect the noise distribution after multiplication. 

For EMCCD detectors, the concept of noise factor $F$ is introduced, which represents the relationship between the standard deviation of the input signal $\sigma_{\text{in}}$, and that of the output signal $\sigma_{\text{out}}$. 
\begin{equation}
 F = \frac{\sigma_{\text{out}}}{\sigma_{\text{in}} g_A}
\end{equation}

According to \cite{robbins03}, the noise factor takes the following form:
\begin{equation}
F^2 = \frac{1}{g_A} + 2(g_A - 1)g_A^{-\left(\frac{N + 1}{N}\right)}
\end{equation}
It can be seen that when the EM gain is large, $F$ is approximately equal to $\sqrt{2}$, and when $g_A = 1$, $F = 1$. 
Our simulation program can provide an accurate simulation of noise characteristics, which is crucial for the development of camera calibration procedures. In this calibration, the determination of parameters $F$, as well as the photon transfer curve (PTC), system gain, and EM gain all depend on the noise distribution. Therefore, ensuring the accuracy of simulation results is of great importance for the verification of subsequent camera calibration procedures. 

\subsubsection{EM gain - voltage relationship}
The multiplication in an EMCCD is controlled by the voltage applied to the multiplication register. According to \cite{robbins09}, the relationship between EM gain $g_A$ and voltage $V_{EM}$ satisfies the following equation:
\begin{equation}
    g_A = \exp\left[N \beta \exp(\gamma V_{EM})\right],
    \label{eq:em}
\end{equation}
where $\gamma$ is a parameter of the chip itself, related to the design of the multiplication register and the number of registers, $\beta$ is related to temperature and aging effect of the chip. It can be seen that the degree of EM gain is very sensitive to the voltage, with the increase rate surpassing exponential relationships. For the camera hardware, a control parameter (hereafter $E_{in}$) from 1 to 1024 is injected to change the multiplication voltage. Calibrating this relationship is essential in the data reduction pipeline, thus modeling this relationship is necessary for the simulation program.

\begin{figure}
\centering
\includegraphics[width=12cm]{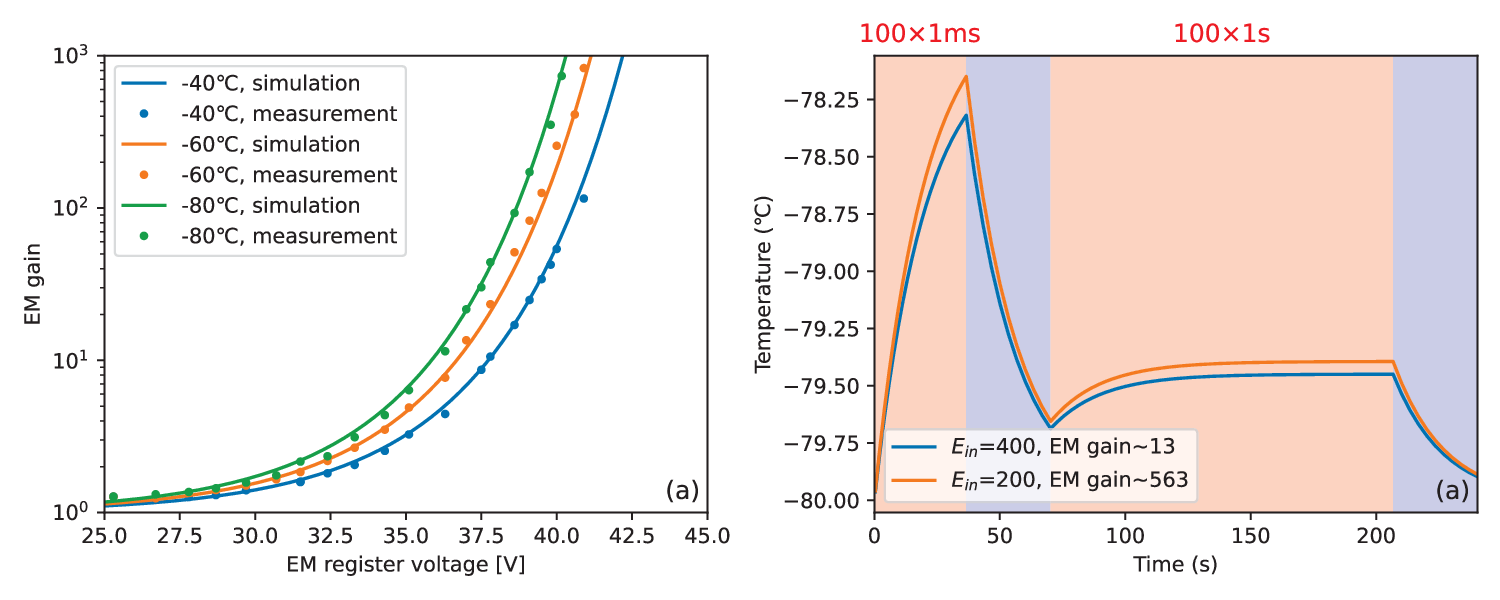}
\caption{(a) Relationship between EM gain and multiplication voltage. The blue, orange, and green lines represent the EM gain model at temperatures of $-40\,^{\circ}\mathrm{C}$, $-60\,^{\circ}\mathrm{C}$, and $-80\,^{\circ}\mathrm{C}$, respectively. The dots indicate the measured results of CPI-C's visual band camera. (b) Example of the camera temperature change curve during sequential exposures consists of 100 frames of 1ms exposuresand 100 frames of 1s exposures.  The red and blue lines represent the cases with $E_M = 200$ (EM voltaEM gain of approximately 563) and $E_M = 400$ (EM gain of approximately 13), respectively. }
\label{em_t}
\end{figure}

The model is built based on the actual testing results of the visible band camera of the CPI-C qualification unit. As the first step in the test, we use an oscilloscope to measure the relationship between \(E_{in}\) and the multiplication voltage \(V_{EM}\). We found that the two variables exhibit a linear relationship. Using the \texttt{polyfit} function from the NumPy package with the parameter \texttt{deg=1}, we fitted the relationship between \(V_{\text{EM}}\) and \(E_{\text{in}}\) as: $V_{EM} = \SI{-1.828e-2}{V} \times E_{in} + \SI{43.61}{V}$.
 The second step is to cool the camera to different temperatures and calibrate the EM gain $g_A$ versus $V_{EM}$ by imaging a point source of the same intensity. We employed the \texttt{curve\_fit} function from the SciPy library to perform data fitting based on Equation \ref{eq:em}, and  obtained the values of \(\beta\) and \(\gamma\). The fitting result shows that \(\gamma=0.2465\) for all temperatures ($T$), and \(\beta = −4.790 \times 10^{-9} \, \frac{T}{1^\circ C} + 1.330 \times 10^{-7}\). In this way, a model of the relationship between electron multiplication and voltage at different temperatures is obtained. Panel (a) of Figure \ref{em_t}a shows the relationship between voltage and multiplication gain at different temperatures. The blue, orange, and green lines represent the EM gain model at temperatures of $-40\,^{\circ}\mathrm{C}$, $-60\,^{\circ}\mathrm{C}$, and $-80\,^{\circ}\mathrm{C}$, respectively. The dots indicate the measured results of the CPI-C visible band camera. It can be seen that the model and measurement fit each other well. As the temperature decreases, the same voltage corresponds to a higher EM gain, and the trend is consistent with the feature of EMCCD.\footnote{}

$\sim$-79.25$^\circ$C
In the testing, we also found that during continuous rapid high EM gain exposure, the temperature of the chip would rise, and the actual EM gain would decrease under the same $E_{in}$. This phenomenon may need to be calibrated and corrected in the pipeline, thus we have added a simple temperature model of the camera to the simulation program. When imaging, the camera temperature will rise slightly, with the rate of temperature increase related to the EM voltage and readout frame rate. After exposure is stopped, the camera temperature decreases according to Newton's law of cooling. Figure \ref{em_t}b shows an example of the camera temperature change curve under the influence of this temperature model. The red curve shows the camera temperature evolution during sequential exposures at $E_M = 200$ (EM gain $\sim$563). The initial temperature of the camera is -80$^\circ$C. The sequence begins with 100 frames of 1ms exposures (total duration $\sim$40s since the readout overhead is 0.365s), causing a temperature rise of $\sim$1.7$^\circ$C. After a 30-second pause where the temperature drops to $\sim$-79.25$^\circ$C. Then the camera takes 100 frames of 1s exposures. During this slower exposure phase, the temperature increases by only $\sim$0.25$^\circ$C. The blue curve displays the identical exposure sequence at $E_M = 400$ (EM gain $\sim$13), which has similar trends but lower peak temperatures due to less heating from the lower EM voltage.

It should be noted that the model related to this effect is a qualitative model and has not been calibrated with actual measurements. The main purpose is to assess the impact of such effects on calibration data processing. If this effect does not need to be added in the simulation, it can be turned off by changing the relevant parameters in the configuration file.

\subsection{Readout Process Modeling}
When a light signal passes through the EMCCD imaging chip, it is converted into an electrical signal, which is then amplified by the front end, converted by an AD converter, and read out as a digital signal. The coronagraph visible light imaging camera uses the LM98640 chip for analog-to-digital conversion (ADC). EMCCD readout modeling includes several parts such as ADC, vertical striping, interference signals, background and background drift, and readout noise.

It should be noted that the camera's interference is added based on the images tested when the visible light camera was not integrated, and there will be differences in the integrated state, which can be adjusted after the final qualification unit test of CPI-C.

\begin{enumerate}
\item Prescan and Overscan. Sixteen non-imaging columns (8 on each side) are read from the left and right edges of the sensor, outside the physical imaging area. These regions are vertical prescan and overscan columns, and can be used for readout noise and bias offset correction. The final image dimensions are 1088 (width) $\times$ 1050 (height) pixels, including these reference columns.

\item AD Conversion. According to actual measurements, the coronagraph ADC module converts 59 electrons after electron multiplication into one ADU, i.e., the system conversion gain $G_{s}$ at $g_A=1$ is 59e$^-$/ ADU.

\item Vertical Striping. The LM98640 chip has two readout taps that alternately perform readout. Due to slight differences between the two taps, vertical stripes will form on the image.

\item Interference. During the camera's readout process, random horizontal stripes appear in the image due to interference from the ground line. We measure the power spectrum of these random horizontal stripes from test images and generate simulated stripes based on the power spectrum and a random phase. 

\item Based on the test results, we observed that the bias level of the camera drifts during imaging. Specifically, a random drift with $\sigma \approx 3$ ADU is observed. Additionally, the bias level decreases when using high EM gain. Both of these effects are modeled in \texttt{CPISM}.

\item Finally, in the simulation, we add Gaussian-distributed readout noise to the image.
\end{enumerate}

\subsection{Example Images}
\begin{figure}
\centering
\includegraphics[width=12cm]{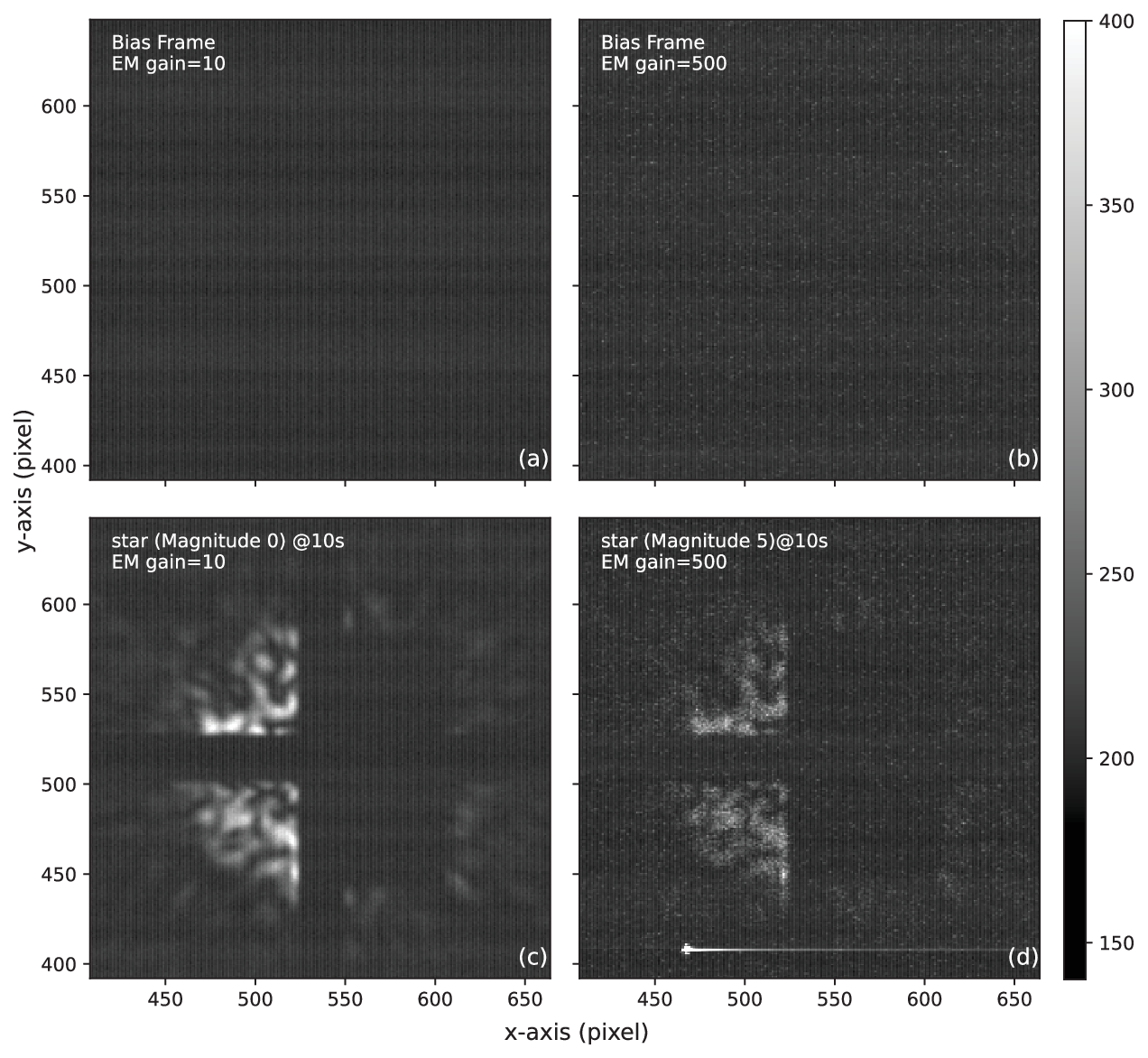}
\caption{Examples of simulated images of CPI-C's visible band camera. (a) Bias image with a multiplication gain of 10. (b) Bias image with EM gain 500. (c) 10-second exposure image of a 0th-magnitude star. (d) 10-second exposure image of a 5th-magnitude star with EM gain 500.}
\label{example}
\end{figure}

In Figure \ref{example}, we present examples of four simulated images. Subfigure (a) shows a zero-exposure bias frame with a multiplication gain of 10, primarily reflecting the readout characteristics of the camera. Horizontal and vertical stripes can be clearly seen. Subfigure (b) displays a bias frame with an EM gain of 500, where the clock-induced charge (CIC) is amplified by the EM gain, resulting in scattered bright spots across the image. In Subfigure (c), we illustrate a 10-second exposure image of a 0th-magnitude star observed by the system. The diffraction pattern of the star and the $10^{-8}$ high-contrast dark holes  are clearly observed. Subfigure (d) demonstrates the imaging of a 5th-magnitude star under the same 10-second exposure but with an increased EM gain of 500. This enhanced multiplication gain improves the detection capability for faint targets. Additionally, a cosmic ray event can be observed in the image. This charge caused by a cosmic ray is also amplified by the EM multiplication process and exceeds the full-well capacity of the multiplication register, causing horizontal trailing in the image. During data reduction, the trails can be partially removed by fitting their profile. However, the inherent photon noise within the trail itself cannot be eliminated. If a trail falls within the dark holes, it may degrade the signal-to-noise ratio (SNR) of the target. To reduce the impact of cosmic rays, careful planning of observations is needed. While long frame exposure times and high EM gain reduce the impact of readout noise, they simultaneously increase the number of cosmic ray events in each frame and raise the probability of cosmic rays exceeding the full-well capacity of the EM register. Thus each observation run demands optimization of exposure time and EM gain settings, and the \texttt{CPISM} program has potential to be used to verify and evaluate the optimization.

\section{Summary}
CPI-C is a specialized high-contrast imaging instrument for exoplanet studies on the CSST. To support the development of its data reduction system and the evaluation of its scientific performance, we have developed an image simulation program \texttt{CPISM}. In \cite{zhu24}, we introduced the overall simulation program and the observation target simulation module, and presented a preliminary analysis of the CPI-C's scientific observation capabilities. This paper is the second in the series, in which we introduce the simulation of the coronagraph instrument, focusing on two core modules: the high-contrast imaging module and the EMCCD camera module.

For the high-contrast imaging module, based on the HCIPy package, we modeled the core high-contrast imaging components such as the pupil apodization filter, the wavefront corrector, and the focal plane mask. We also considered the effects of a wide spectral band, and generated a high-contrast imaging dark zone that is similar to the measured morphology and meets the technical requirements. During the optimization of the high-contrast imaging dark zone, we explored the use of electric field conjugation methods for dark zone contrast optimization, which is also the first application of this method in pupil transmission modulation high-contrast systems similar to CPI-C.

For the EMCCD module, we modeled various effects during the image collection process, transfer process, EM multiplication process, and readout process, achieving simulation images that are very close to the actual instrument images. These simulation images contain rich details, providing input for the development of the CPI-C data reduction pipeline. In the simulation of the EMCCD, to accelerate the modeling of the EM multiplication process, we proposed a more accurate formula for the EMCCD multiplication noise distribution. This formula can also be applied to fields such as EM gain fitting.

\normalem
\begin{acknowledgements}
This work is supported by the project of the CSST scientific data processing and analysis system of the China Manned Space Project. We thank the anonymous referee for valuable comments that have significantly improved the manuscript. This work is supported by the National Natural Science Foundation of China (NSFC) under grant nos U2031210 and 11827804, as well as the science research grants from the China Manned Space Project (CMS-CSST-2021-A11, CMS-CSST-2021-B04, CMS-CSST-2025-A17, CMS-CSST-2025-A18, CMS-CSST-2025-A19) and the Pre-research project on Civil Aerospace Technologies funded by China National Space Administration (No. D010301).
\end{acknowledgements}

\appendix
\section{Electric Field Conjugation Optimization Method} 

The core objective of the Electric Field Conjugation (EFC) optimization is to iteratively minimize residual starlight within a designated region of the focal plane, forming a high-contrast "dark hole" (DH). The EFC process in the \texttt{CPISM} simulation follows \citep{matthews17a} and the document of the HCIPy package. \footnote{ https://docs.hcipy.org/0.6.0/tutorials/ElectricFieldConjugation/ElectricFieldConjugation.html}.

Consistent with the CPI-C design parameters, the target dark hole for optimization was defined as a single-sided region extending from an inner working angle (IWA) of 3 $\lambda/D$ to an outer working angle (OWA) of 16 $\lambda/D$. 
The simulation models the main optical path of CPI-C, including modeled system aberrations, the transmission apodizing filter, the wavefront corrector, up to the final focal plane. Crucially, instead of simulating detector readout and noise, the optimization process directly accesses the complex electric field data generated at the focal plane.

As the first step of EFC optimization, we calculate the Jacobian matrix $G$ of the system, which links the command voltages applied to the wavefront corrector's actuators to the resulting complex electric field within the DH. We numerically simulate the "poke" method, in which we sequentially apply small, calibrated perturbations ($\pm \delta u$) to each individual actuator of the DM model and record the corresponding change in the complex electric field vector $E_{\text{DH}}$ within the designated DH region. The \(i\)-th column of the Jacobian has the form of:
\begin{equation}
    \mathbf{g}_i = \frac{\partial E_{\text{DH}}}{\partial u_i} \approx \frac{E_{\text{DH}}(u_i + \delta u) - E_{\text{DH}}(u_i - \delta u)}{2 \delta u}
    \label{eq:jacobian_poke}
\end{equation}
where $E_{\text{DH}}(u_i \pm \delta u)$ denotes the complex electric field vector in the DH when only the \(i\)-th actuator's command is perturbed by $\pm \delta u$ from a base state.

% Regularization
As the Jacobian matrix $G$ can be ill-conditioned or non-square, directly inverting it to find the correction commands is often numerically unstable. To address this, Tikhonov regularization is employed to compute a stable, regularized pseudo-inverse $G^\dagger_{\lambda}$ of the Jacobian.

After finding the Jacobian matrix, we optimize the dark hole by iteration. For each iteration $k$:
\begin{enumerate}
    \item The current complex electric field $E_k$ within the DH is obtained. As mentioned, in this simulation, this step is simplified by directly accessing the field $E_k$ from the optical propagation model.
    \item Based on the measured field $E_k$ and the regularized pseudo-inverse $G^\dagger_{\lambda}$, a correction voltage command $\Delta u_k$ for the wavefront corrector is calculated as:
    \begin{equation}
        \Delta u_k = -\gamma G^\dagger_{\lambda} E_k
        \label{eq:efc_control_law}
    \end{equation}
    where $\gamma$ is the loop gain ($0 < \gamma \le 1$)  to moderate the correction step size.
    \item The total voltage command vector applied to the wavefront corrector is updated for the next iteration: $u_{k+1} = u_k + \Delta u_k$.
\end{enumerate}
This iterative process continues until a desired contrast level in the DH is achieved or a maximum number of iterations is reached.

In summary, to generate a dark hole in simulation with characteristics closely resembling the design goals of CPI-C, we utilized the EFC method for focal plane wavefront control. The resulting optimized dark hole spans the 3-16 $\lambda/D$ region and achieves a contrast level of approximately \(10^{-8}\), consistent with the instrument's target performance. Since EFC is a model-based optimization technique, EFC takes advantage of the simulation where the system model is explicitly defined. Furthermore, the simulation environment allows direct access to the complex electric field, greatly simplifying the sensing step compared to real optimization, which requires sophisticated estimation techniques. 
Implementing EFC on the actual instrument would be a challenge related to model accuracy, calibration, and real-time electric field estimation from detector images. In future work, we will explore the potential use of EFC in the transmission apodization coronagraph system, such as the next generation of CPI-C.

\section{Parameters of CPI-C instrument}
In Table~\ref{parameters}, we list key parameters for the CPI-C instrument currently configured in our simulation program. It should be noted that most parameters are currently set according to the instrument's design specifications or requirement values. In future software versions, we will update these parameters based on ground-based or in-orbit measured values.
\begin{table}[h]
\centering
\caption{Parameters of CPI-C instrument used in CPISM}
\setlength{\tabcolsep}{10pt}
\label{parameters}
\begin{tabular}{cc}
\hline\noalign{\smallskip}
Parameter Name & Parameter Value or Range \\
\hline\noalign{\smallskip}
Dark zone contrast & \num{1E-08} \\
Inner working angle (IWA) & \SI{3}{\lambda/D}; \SI{0.205}{\arcsecond}@\SI{662}{\nm} \\
Outer working angle (OWA) & \SI{16}{\lambda/D}; \SI{1.09}{\arcsecond}@\SI{662}{\nm} \\
Width of the cross-shape focal mask & \SI{0.4}{\arcsecond} \\
Focal plane scale & \SI{1.615e-2}{\arcsecond/pix} \\
Camera effective pixels & \num{1024} $\times$ \num{1024}\\
Field of view (FOV) & \SI{16.54}{\arcsecond} $\times$ \SI{16.54}{\arcsecond} \\
Read noise ($1\times$ EM gain) & \SI{160}{e^{-}/pix} \\
Clock induced charge (CIC) & \SI{0.2}{e^{-}/pix} \\
Dark current & \SI{1.0E-03}{e^{-}/pix/s} \\
Conversion gain ($1\times$ EM gain) & \SI{59}{e^{-}/ADU} \\
Full-well capacity of image region & \SI{80000}{e^{-}} \\
Charge handling capacity of multiplication register & \SI{500000}{e^{-}} \\
Readout time & \SI{0.365}{s} \\
\hline\noalign{\smallskip}
\end{tabular}
\end{table}

\section{List of Acronyms}
In Table~\ref{tab:acronyms}, we list all of the acronyms in the paper for reference.
\begin{table}[h]
\centering
\caption{List of Acronyms and Technical Terms}
\label{tab:acronyms}
\begin{tabular}{@{}ll@{}}
\hline\noalign{\smallskip}
\textbf{Acronym} & \textbf{Full Term} \\
\hline\noalign{\smallskip}
CPI-C & Cool Planets Imaging Coronagraph \\
CSST & China Space Station Telescope \\
CPISM & CPI-C Image Simulator \\
CGI & Coronagraph Instrument \\
JWST & James Webb Space Telescope \\
\hline\noalign{\smallskip}
PSF & Point Spread Function \\
IWA & Inner Working Angle \\
OWA & Outer Working Angle \\
EFC & Electric Field Conjugation \\
SPGD & Stochastic Parallel Gradient Descent \\
FOV & Field of View \\
SED & Spectral Energy Distribution \\
PTC & Photon Transfer Curve \\
ADI & Angular Differential Imaging \\
LOCI & Locally Optimized Combination of Images \\
OIRS & Optimized Image Rotation and Subtraction \\
\hline\noalign{\smallskip}
CCD & Charge-Coupled Device \\
EMCCD & Electron Multiplying Charge-Coupled Device \\
CIC & Clock Induced Charge \\
CTI & Charge Transfer Inefficiency\\
ADC & Analog-to-Digital Conversion\\
MEMS & Micro-electromechanical Systems \\
DM & Deformable Mirror \\
\hline\noalign{\smallskip}
\end{tabular}
\end{table}

\bibliographystyle{raa}
\bibliography{bibtex}

\label{lastpage}

\end{document}